\documentclass[preprint,12pt]{elsarticle}

\usepackage{amssymb}

\usepackage{lineno}

\usepackage[dvipsnames]{xcolor}
\newcommand{\specialcell}[2][c]{%
  \begin{tabular}[#1]{@{}c@{}}#2\end{tabular}}

\journal{Nuclear Physics A}

\begin{document}
\begin{frontmatter}



\title{Concept for a Space-based Near-Solar Neutrino Detector}

\author[inst1]{N. Solomey\fnref{nSolo}}
\fntext[nSolo]{\textit{Email:} nick.solomey@wichita.edu}
\author[inst1]{J. Folkerts\fnref{jFolk}\corref{cor1}}
\fntext[jFolk]{\textit{Email:} jdfolkerts@shockers.wichita.edu \textit{Mailing address:} Jonathan Folkerts, Wichita State University, Campus box 32, 1845 Fairmount St., Wichita, KS 67260 \textit{Telephone:} (316) 978-3991}
\author[inst1]{H. Meyer\fnref{hMeye}}
\fntext[hMeye]{\textit{Email:} holger.meyer@wichita.edu}
\author[inst1]{C. Gimar}
\author[inst1]{J. Novak\fnref{jNova}}
\fntext[jNova]{\textit{Email:} jcnovak@shockers.wichita.edu}
\author[inst1]{B. Doty\fnref{bDoty}}
\fntext[bDoty]{\textit{Email:} bsdoty@shockers.wichita.edu}
\author[inst1]{T. English\fnref{tEngl}}
\fntext[tEngl]{\textit{Email:} txenglish@shockers.wichita.edu}
\author[inst1]{L. Buchele}
\author[inst1]{A. Nelsen}

\author[inst2]{R. McTaggart\fnref{rMcTa}}
\fntext[rMcTa]{\textit{Email:} robert.mctaggart@sdstate.edu}

\author[inst3]{M. Christl\fnref{mChri}}
\fntext[mChri]{\textit{Email:} mark.christl@nasa.gov}
\cortext[cor1]{Corresponding Author}


\affiliation[inst1]{organization={Wichita State University},
            addressline={1845 Fairmount St.}, 
            city={Wichita},
            postcode={67260}, 
            state={Kansas},
            country={United States of America}}

\affiliation[inst2]{organization={South Dakota State University},
            addressline={175 Medary Avenue}, 
            city={Brookings},
            postcode={57006}, 
            state={South Dakota},
            country={United States of America}}

\affiliation[inst3]{organization={NASA Marshall Space Flight Center},
            addressline={Martin Rd SW}, 
            city={Huntsville},
            postcode={35808}, 
            state={Alabama},
            country={United States of America}}


\begin{abstract}
The concept of putting a neutrino detector in close orbit of the Sun has been unexplored until very recently. The primary scientific return is to vastly enhance our understanding of the solar interior, which is a major NASA goal. Preliminary calculations show that such a spacecraft, if properly shielded, can operate in space environments while taking data from neutrino interactions. These interactions can be distinguished from random background rates of solar electromagnetic emissions, galactic charged cosmic-rays, and gamma-rays by using a double pulsed signature. Early simulations of this project have shown this veto schemes to be successful in eliminating background and identifying the neutrino interaction signal in upwards of 75\% of gamma ray interactions and nearly 100\% of other interactions. Hence, we propose a new instrument to explore and study the Sun. Due to inverse square scaling, this instrument has the potential to outperform Earth-based experiments in several domains such as making measurements not accessible from the Earth's orbit.

\end{abstract}



\begin{highlights}
\item It may be possible to build a neutrino detector for close approach to the Sun.
\item Neutrino flux increases by 10,000x at $3R_\odot$ increasing effective mass for close approach.
\item Double-coincidence gallium neutrino signals allow for rejecting high-rate space backgrounds.
\end{highlights}


\end{frontmatter}


\section{Introduction}\label{sec:Introduction}
Since neutrinos only interact weakly, they are hard to detect; nevertheless, within the last ten years neutrino detectors on Earth have started to reliably detect neutrinos from the fusion reactions inside the Sun and scientists have started to use this information to investigate the Sun’s nuclear furnace. For example, Super Kamiokande (Super K) has been able to produce a now-famous image of the sun in Neutrinos\cite{ref:SuperKDetector}, and Borexino has been able to experimentally verify high metallicity standard solar models' (SSM) predicted neutrino fluxes\cite{ref:BorexinoSolar}. These neutrino detectors are normally very large and deep underground. They are large because they need as much mass as possible to capture more neutrinos, and they are underground to reduce background rates. Clearly there is a scientific advantage associated with the increased neutrino flux of flying near the sun, but there is also scientific merit in designing a space-capable detector which can get further from the sun and benefit from the decreased solar neutrino flux to search for galactic neutrinos or for weakly interacting dark matter candidates.	

We propose three distinct possibilities for science using the Neutrino Solar Observatory ($\nu$SOL):
\begin{enumerate}
    \item Going closer to the Sun, see Table \ref{tab:fluxTable}, the $1/r^2$ neutrino flux would provide 1,000x more neutrinos per second at a distance of seven solar radii ($7 R_\odot\approx 5\cdot10^6$ km), approximately where the NASA Parker Solar Probe currently operates. Getting as close as 3 $R_\odot$, which some NASA scientists think is possible, the neutrino flux would be increased by a factor of 10,000x \cite{ref:StudyingFurnaceNeutrinos}. In this regime, there are several opportunities for new science. First, such a detector may be able to make statistical measurements better than currently possible. Such a detector could perform an analysis similar to one performed by Borexino to verify predictions of the SSM \cite{ref:BorexinoSolar}. By flying in the region between the earth ($215R_\odot$) and the sun, a space-based detector may be able to search for evidence that confirms the standard predictions of neutrino decoherence by the Sun’s surface, or it may be able to find that the neutrinos are not yet fully de-coherent in regions near the sun. Both of these objectives would be accomplished by looking for deviation from the expected electron-neutrino probability in the measured neutrino flux. Finally, a detector near the sun may be able to indirectly search for dark matter by looking for disruptions to the fusion region inside the sun \cite{ref:StudyingFurnaceNeutrinos}.
    \item Because the neutrino has a non-zero mass, which we know from the existence of neutrino oscillations, the Sun bends space to create a gravitational focus. For neutrinos, this gravitational focus would be very close to us at 20 to 40 AU ($3.0\cdot10^{9}$ to $6.0\cdot10^{9}$ km). This distance is easily reachable with current technology, unlike the gravitational focus for light which is 450 to 750 AU ($6.7\cdot10^{10}$ to $1.1\cdot10^{11}$ km) away \cite{ref:GravFocus}. This solar neutrino focus could be a testing ground for gravitational lens ideas. The galactic core is 27,000 light years away from us, is about two times larger than the moon when viewed from earth, and is the 2nd largest neutrino source in the sky after the sun \cite{ref:GalacticCoreDistance,ref:CoreBlackHoles,ref:CoreNeutronStars}. The galactic core not only has many neutrino-producing stars, but it also $\sim10,000$ neutron stars and black holes in the central cubic parsec of the core. Matter falling into these objects is crushed, producing neutrons and isotropically emitted neutrinos of higher energy than solar fusion neutrinos. A detector at the solar neutrino focus would permit imaging of the galactic core and just finding the neutrino gravitational focus of the Sun would be a new way to measure the neutrino mass. Such a gravitational focus mission would be more limited in terms of studying fusion, but would be in a lower radiation environment than a mission to the sun.
    \item A space-based neutrino probe can take advantage of the changing shape of the $1/r^2$ neutrino flux along highly elliptical orbit; when we approach the sun, we expect to see a curve matching the $1/r^2$ modulation of the SSM neutrino emission. When traveling away from the Sun, deviations from the expected $1/r^2$ curve are an indication of the direct observation of dark matter or galactic neutrinos \cite{ref:StudyingFurnaceNeutrinos}.
\end{enumerate}

Although operating a neutrino detector in space will be challenging, we believe we have found a way to do this. Our technique would permit operation and detection in space, would be a major advance for astrophysics and heliophysics, and would allow for new ways to make elementary particle physics fundamental measurements, all of great importance.

\begin{table}[htbp]
    \centering
    \begin{tabular}{lrl}
        \textbf{Distance from Sun} & \textbf{Flux relati}&\hspace{-4.2mm}\textbf{ve to Earth} \\\hline\hline
        696342 km ($R_\odot$)& 46200&\\\hline
        1500000 km ($\sim3R_\odot$) & 10000&\\\hline
        4700000 km ($\sim7R_\odot$) & 1000&\\\hline
        15000000 km & 100&\\\hline
        47434000 km & 10&\\\hline
        Mercury & 6&\hspace{-4.2mm}.4\\\hline
        Venus & 1&\hspace{-4.2mm}.9\\\hline
        Earth & 1&\hspace{-4.2mm}(6.544$\cdot10^{10}\textrm{ cm}^{-2}\textrm{ s}^{-1}$)\\\hline
        Mars & 0&\hspace{-4.2mm}.4\\\hline
        Asteroid Belt & 0&\hspace{-4.2mm}.1\\\hline
        Jupiter & 0&\hspace{-4.2mm}.037\\\hline
        Saturn & 0&\hspace{-4.2mm}.011\\\hline
        Uranus & 0&\hspace{-4.2mm}.0027\\\hline
        Neptune & 0&\hspace{-4.2mm}.00111\\\hline
        Pluto & 0&\hspace{-4.2mm}.00064\\\hline
        KSP & 0&\hspace{-4.2mm}.0002\\\hline
        Voyager 1 (2015) & 0&\hspace{-4.2mm}.00006\\\hline
    \end{tabular}
    \caption{Intensity of solar neutrinos at various distances from the Sun relative to the flux at Earth and total solar neutrino flux at earth\cite{ref:Bahcall_2001}.}
    \label{tab:fluxTable}
\end{table}

\section{Detector Design}\label{sec:Design}
$\nu$SOL is designed around an explorer-class NASA Medium-Class Explorers (MIDEX) mission. Within this regime, the satellite would operate as a technical demonstrator mission with the goal of an operational proof-of-concept by measuring neutrinos in space and allowing for a flagship mission following successful demonstration. In keeping with this class of mission, we compare some components of the two designs that our project has used to date with the characteristics of a similarly priced technical demonstrator, Energetic Gamma Ray Experiment Telescope (EGRET), which flew on the Compton Gamma Ray Observatory (CGRO) \cite{ref:CGRO-EGRET}.

The science payload of this detector consists of an inner science detector surrounded by an active vetoing system, which is in turn surrounded by a passive shield. The science payload has changed the most since the detector’s conception, and it will likely be the component that continues to change the most radically as design improvements or alternate designs are explored. The active vetoing system has consistently remained a plastic scintillating volume encapsulating the central science detector. The outer passive shielding has undergone some simple design refinements throughout the process. In both the initial design and first revision, the detector is a right-circular cylinder to maximize the volume behind the heat shielding.

All versions of the detector's designs made so far have used Gallium as a neutrino target. This choice has been made for several reasons, which are elaborated on in Section \ref{sec:GaDoublePulse}. What is relevant to the design here is that gallium provides a double-pulse signal from the prompt electron and the nuclear de-excitation $\gamma$-ray within hundreds of nanoseconds. The fast timing allows for highly accurate rejection of single-pulsed background signals. This interaction has the drawback in that it is only sensitive to electron-type neutrinos, and not the mu- and tau-types.
\subsection{Initial Design}\label{sec:Design:FirstDesign}
\begin{figure}[htbp]
    \centering
    \includegraphics[width=0.95 \textwidth]{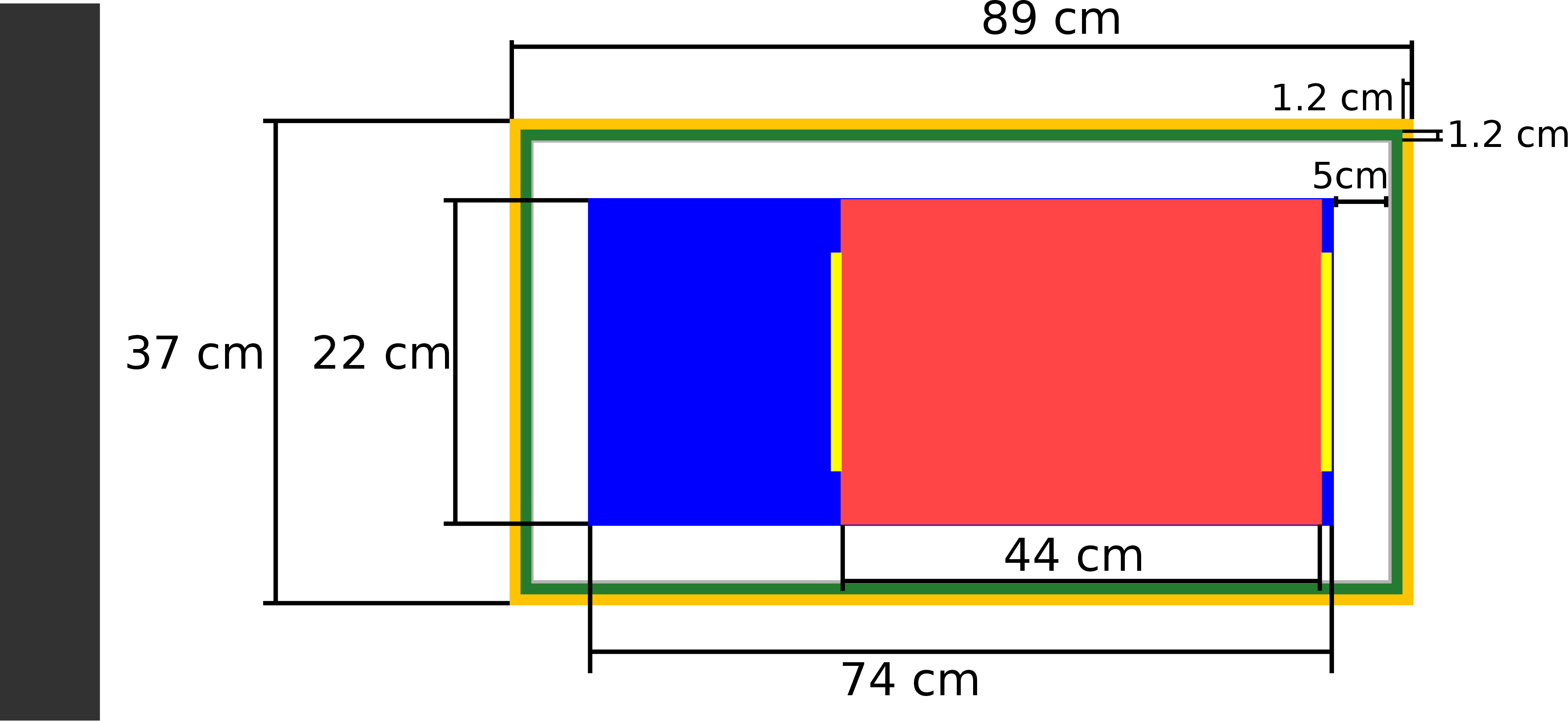}
    \caption{Image of first detector science payload design. Sunward direction is to the left with the solar EM and heat shielding (dark grey, not to scale). From outside in, the iron shell (orange), copper-polymer showering layers (green), Plastic showering layer (light grey), and vetoing volume (white) all surround the inner volume. This inner volume consists of a volume of liquid scintillator (red) capped by two PMTs (yellow) which are both surrounded by a reflector and space for control electronics (blue).}
    \label{fig:DetctorV0}
\end{figure}

An image of most of the detector design is shown in Figure \ref{fig:DetctorV0}.
The first layer of shielding in the first design is two separate pieces of metal for passive radiation shielding. The entire detector is surrounded by a 1 cm thick layer of iron, and a 10 cm thick cylindrical tungsten plate sits on the Sunward side of the main volume to provide shielding from the solar wind charged particles. Iron was chosen as the passive shield for all shielding needs near the science volume because of its ease of fabrication, commonplace use, and resistance to becoming radioactive in high-radiation environments. Beneath the iron shielding is a 1 cm thick polymer shell followed by several 0.2 cm thick alternating layers of aluminum and plastic to promote showering of any particles that penetrate the passive iron shield.

\begin{table}[htbp]
    \centering
    {\small
    \begin{tabular}{|l|c|c|}\hline
       Parameter  & EGRET & $\nu$SOL Design 1 \\\hline\hline
     Detector Type:  &\specialcell{Spark chambers,\\NaI(Tl) crystals,\\and plastic scintillators} & \specialcell{Doped liquid scintillator\\and plastic scintillators}\\\hline
Energy Range: & \specialcell{20 MeV to about\\30 GeV} & 0.235 to $\sim20$ MeV\\\hline
Energy Resolution: & \specialcell{Approximately twenty\\percent over the\\central part of\\the energy range.} & $\sim 40\%$\\\hline
Total Detector Area: & Approximately 6400 cm$^2$ & $\sim 100 $ cm$^2$ \\\hline
Effective Area: & \specialcell{Approximately 1500 cm$^2$\\between 200 MeV and\\1000 MeV, falling at\\higher and lower energies} & $<1.6\cdot 10^{-23}$ cm$^2$ \\\hline
Timing Accuracy: & 0.1 ms absolute & $\sim 10$ ns \\\hline
Weight: & about 1830 kg (4035 lbs) & $\sim365$ kg (800 lbs)\\\hline
Size: & 2.25 m x 1.65 m diameter & \specialcell{0.83 m x 0.45 m\\diameter main payload\\and 0.1 m x 0.60 m\\diameter solar wind shield.} \\\hline
    \end{tabular}
    }
    \caption{Comparison of first $\nu$SOL design with several EGRET parameters.}
    \label{tab:design1_vs_EGRET}
\end{table}

Inside the outer shield and showering layers is a 5 cm thick vetoing volume made of plastic scintillator. The veto scintillator is a single hollow shell to completely encapsulate the inner detector. This volume looks for any energy depositions in coincidence with signals inside the science volume, and sends a rejection signal in the timing window following any deposition inside the veto. Inside this volume is a space reserved for  electronics and also the main detector volume. The science volume consists of a single right-circular cylinder of gallium-doped liquid scintillator (LS). This LS volume has a radius of 15 cm, and a length of 44 cm. The design used an oil-based LS, as in the NO$\nu$A experiment \cite{ref:NOvA_LS}, but the detector is not segmented like NO$\nu$A.

To test this science volume, we acquired a sample of the LS used by NO$\nu$A, and we mixed in several gallium compounds, including Gallium II/III Chloride and Gallium II/III Oxide, to test the solubility of the compounds in oil-based LS. By way of optical spectroscopy, we determined that several of the gallium compounds can cause the wavelength shifters in NO$\nu$A LS to crash out of solution. Using flame atomic absorption (FLAA) and inductively coupled plasma (ICP) spectroscopy, we found no gallium compounds that can have a concentration greater than 50 mg/L in mineral-oil based scintillator. This corresponds to a mass fraction of less than 1/10,000. A dopant fraction this small and the difficulty of keeping necessary wavelength shifters in solution were the primary driving factors to lead our research away from liquid scintillators for a solar mission.

\subsection{1st Revision}\label{sec:Design:SecondDesign}
The first revision of the detector design contains several major changes to the basic design: 
\begin{itemize}
    \item Due to the difficulty of loading liquid scintillator with gallium, we have designed using Cerium-doped Gadolinium-Aluminum-Gallium Garnet (Ce:GAGG), often referred to as simply GAGG. This is a recently developed scintillating crystal which is approximately 22\% gallium by mass. Besides having a high mass-fraction of gallium, GAGG also has a fast decay time (50-150 ns) and a high light yield (40,000-60,000 photons/MeV) varying with different cerium dopant levels.
    \item We have changed from a single large scintillator to a series of four segmented detectors, each with its own pair of photodetectors.
    \item Current designs have removed the showering layers of plastic and aluminum to save mass.
    \item We have redesigned to use silicon photomultipliers (SiPMs). At the temperatures expected for a detector using a heat shield like that on the Parker Solar Probe, the SiPMs would have noise orders of magnitude smaller than the signals from the high-yield GAGG scintillation, on the order of 2000 photons for 50 keV signals.
    \item A ratio of radius to length for the detector was chosen at 1:8 for a closest approach of 10 $R_\odot$ with a safety factor for heat shield pointing. Detector dimensions were chosen based on the largest mass possible within this heat shielding constraint.
\end{itemize}

\begin{table}[htbp]
    \centering
    {\small
    \begin{tabular}{|l|c|c|}\hline
       Parameter  & EGRET & $\nu$SOL Revision 1\\\hline\hline
     Detector Type:  &\specialcell{Spark chambers,\\NaI(Tl) crystals,\\and plastic scintillators} & \specialcell{GAGG scintillator\\and plastic scintillators}\\\hline
Energy Range: & \specialcell{20 MeV to about\\30 GeV} & 0.235 to $\sim20$ MeV\\\hline
Energy Resolution: & \specialcell{Approximately twenty\\percent over the\\central part of\\the energy range.} & $\sim 40\%$\\\hline
Total Detector Area: & Approximately 6400 cm$^2$ & $\sim 100 $ cm$^2$ \\\hline
Effective Area: & \specialcell{Approximately 1500 cm$^2$\\between 200 MeV and\\1000 MeV, falling at\\higher and lower energies} & $\sim 7.5\cdot 10^{-19}$ cm$^2$ \\\hline
Timing Accuracy: & 0.1 ms absolute & $\sim 10$ ns \\\hline
Weight: & about 1830 kg (4035 lbs) & $\sim545$ kg (1200 lbs)\\\hline
Size: & 2.25 m x 1.65 m diameter & \specialcell{1.23 m x 0.16 m\\diameter.} \\\hline
    \end{tabular}
    }
    \caption{Comparison of the current revision of the $\nu$SOL design with several EGRET parameters.}
    \label{tab:design2_vs_EGRET}
\end{table}

\begin{figure}[htbp]
    \centering
    \includegraphics[width=0.95 \textwidth]{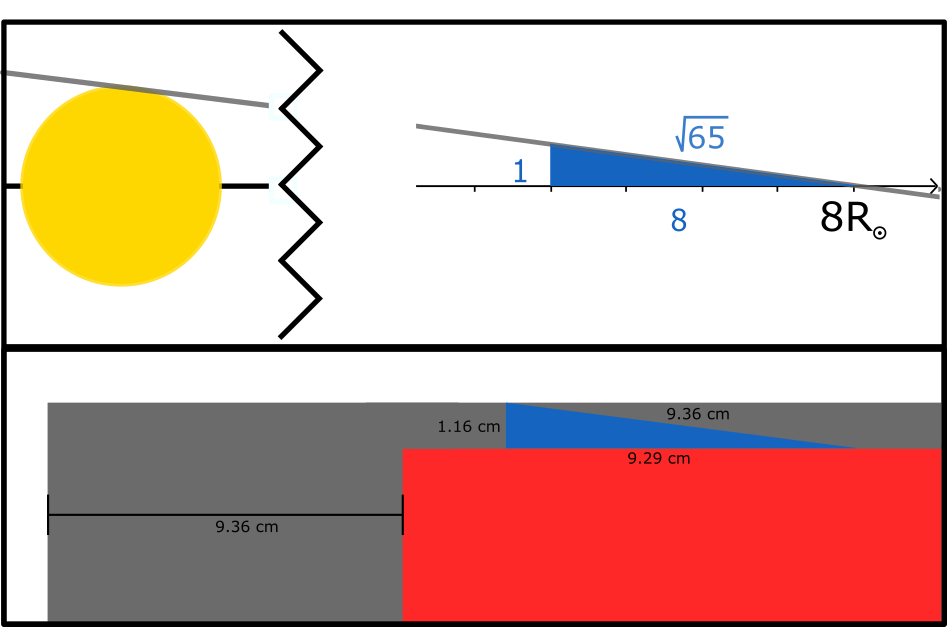}
    \caption{Top: Geometry of solar protons coming from the sun showing the triangle made by a proton from the pole of the sun through a point near the equatorial plane. At a distance of $8R_\odot$, the distance a proton needs to cross is at least $\sqrt{65}$x longer than the height of the triangle. Bottom: Section of the detector with the front-facing shield and the triangle from above overlaid on the side of the iron shielding. Notice that a 1.16 cm thick shielding is the equivalent of 9.36 cm for protons coming in at angles allowed by the geometry of the sun.}
    \label{fig:SunGeometry}
\end{figure}

Lastly, the front-facing tungsten shield has been replaced with iron shielding directly abutting the science payload and thickened the iron shell of the detector to 1.17 cm thick. The solar shield now consists of a thick front-facing section of iron with the same diameter as the rest of the science payload to create a total front-facing iron thickness of 9.36 cm. We changed the dimensions of the outer iron shielding because the solar wind protons come in at a sufficiently shallow angle that they experience very long sections of the relatively thin outer shield which are equivalent to having a thick front-facing shield, Figure \ref{fig:SunGeometry}. In this geometry, 1.17cm thickness creates a CSDA track longer than the requirement of 9.36 cm. A thickness of 9.36 cm was chosen based on the continuously slowing down approximation (CSDA) range of protons in iron, and this thickness is sufficient to stop protons up to 350 MeV \cite{ref:NIST_CSDA_Ranges}. The size of this front-facing shield was one of the major drivers of detector mass, and the other was the overall diameter of the detector. By changing to this smaller diameter front-facing shield with slightly thicker iron on the outer shell, the overall mass of iron shield goes down while still stopping a majority of solar protons. This change, coupled with the much denser GAGG material for the central detector has led to orders of magnitude gains in the detector mass relative to the total payload mass.

\begin{figure}[htbp]
    \centering
    \includegraphics[width=0.95 \textwidth]{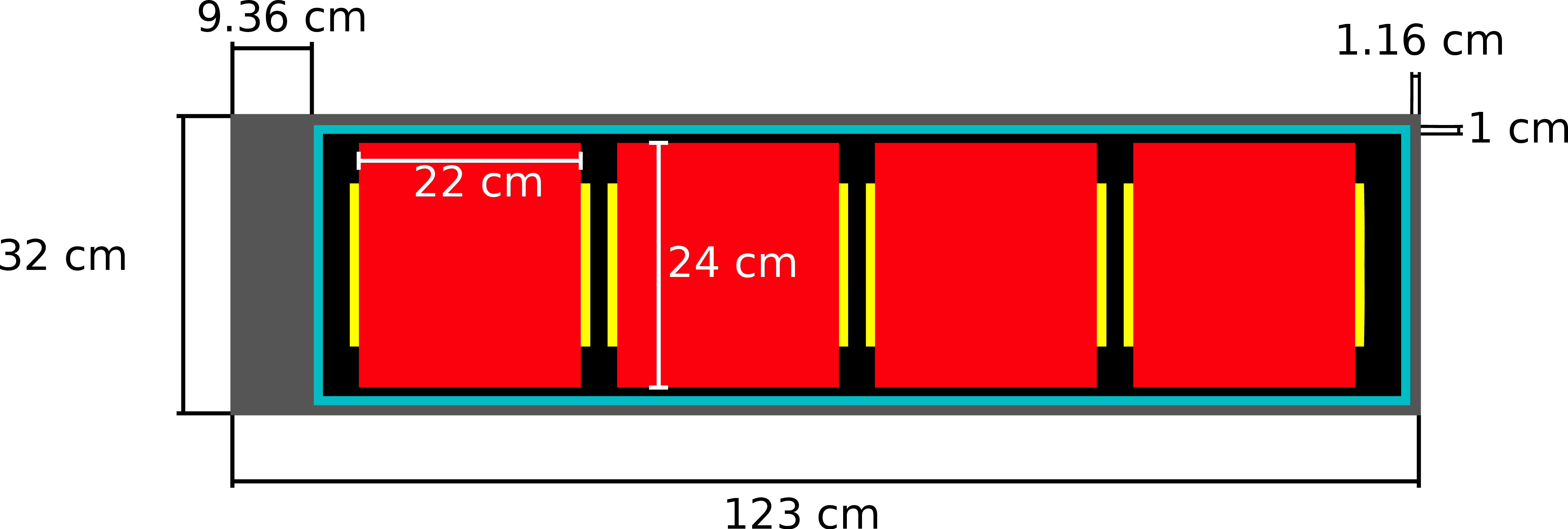}
    \caption{Most recent detector science payload design. A series of four GAGG cylinders (red) are capped with SiPM arrays (yellow) on both ends. Multiple tiled SiPMs would be used to increase the area coverage and to measure in coincidence with one another. These rest inside a cavity for wiring and control electronics (black). This cavity is surrounded by an active veto array (blue) made of scintillating plastic. The outer layer is passive iron shielding (grey) which is thicker on the sun-facing side of the detector.}
    \label{fig:DetctorV1}
\end{figure}

The current design with segmented detectors, thicker iron shielding, and the front-facing shield abutting the detector can be seen in Figure \ref{fig:DetctorV1}.

\subsection{Continuing Design Possibilities}\label{sec:Design:Continuing}
We are at the stage of design where there are still open questions about possible designs for the science volume, and we will present a few concepts that have been considered, but not yet simulated. First, we look to the success of highly segmented detectors such as NO$\nu$A. Nova uses voxels which are oriented in an x-y plane perpendicular to the direction of their beamline. Each voxel runs the width of the detector in one of the x/y directions, a few centimeters in the other direction, and a few centimeters in the z-direction. By alternating the x-oriented and y-oriented layers, NO$\nu$A can tag particles and energies using machine learning algorithms \cite{ref:NOvAEventClassification}. A detector like ours could mimic this sort of design using small GAGG crystals which have a reflective coating and SiPM arrays as endcaps. Crystals on the order of 3mm x 3mm could closely approximate the cylindrical shape important for proper sun shading and provide very fine voxelation. This design has the advantage that there is software that could be modified to parse the data, but there are other challenges that arise, such as the required data throughput.

\begin{figure}
    \centering
    \includegraphics[width=0.4 \textwidth]{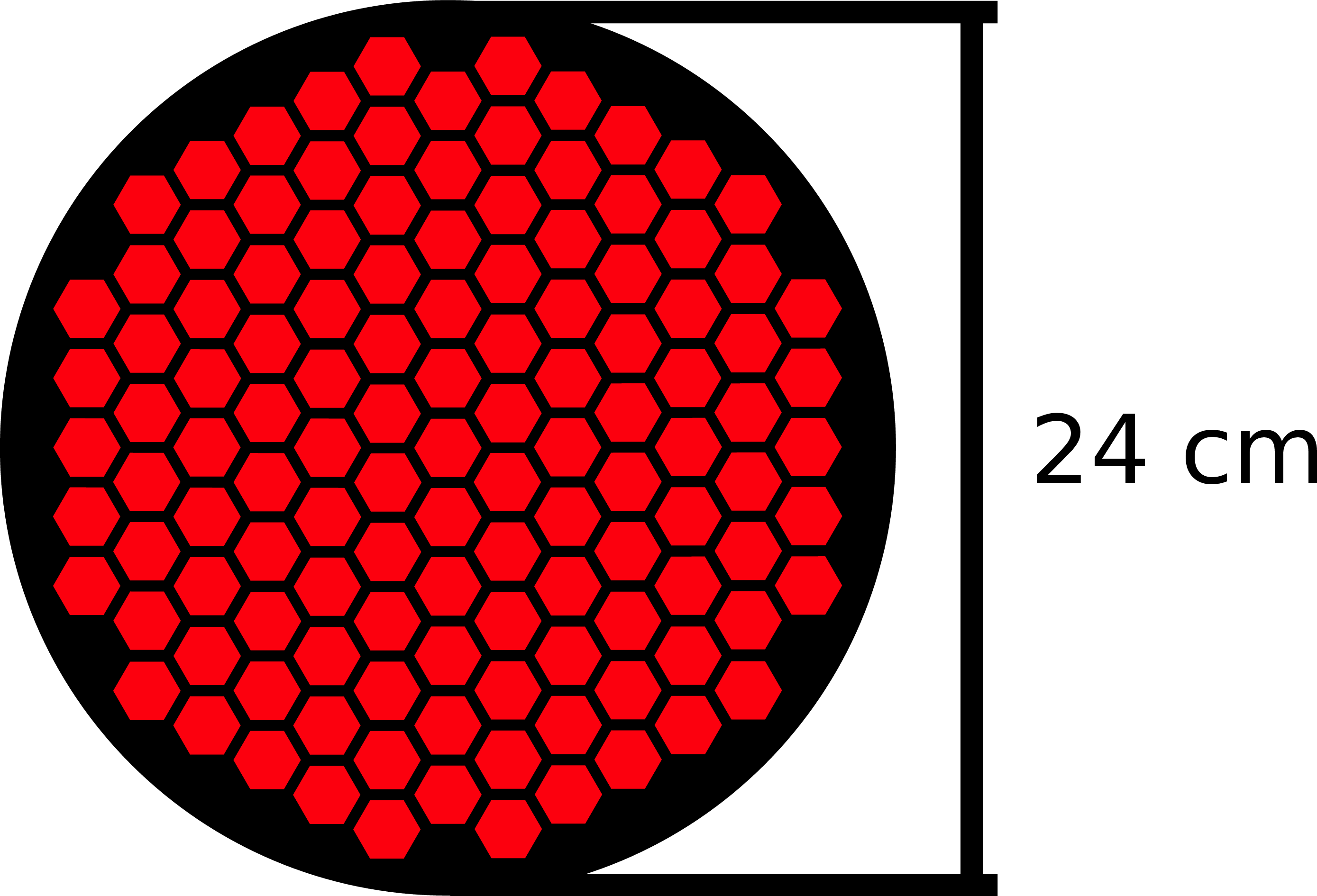}
    \caption{Possible detector design. Sunward direction is into/out of the page. Hexagonal prisms of GAGG (red) fill as much space as possible while leaving room for wires inside a circular shadow cast by the heat shield. As the hexagons become finer, less space is lost from the edges of the circular region, but more space is lost to wiring and SiPMs. Specific placement of SiPMs for this design are not finalized.}
    \label{fig:HexDetector}
\end{figure}

\begin{figure}
    \centering
    \includegraphics[width=0.4 \textwidth]{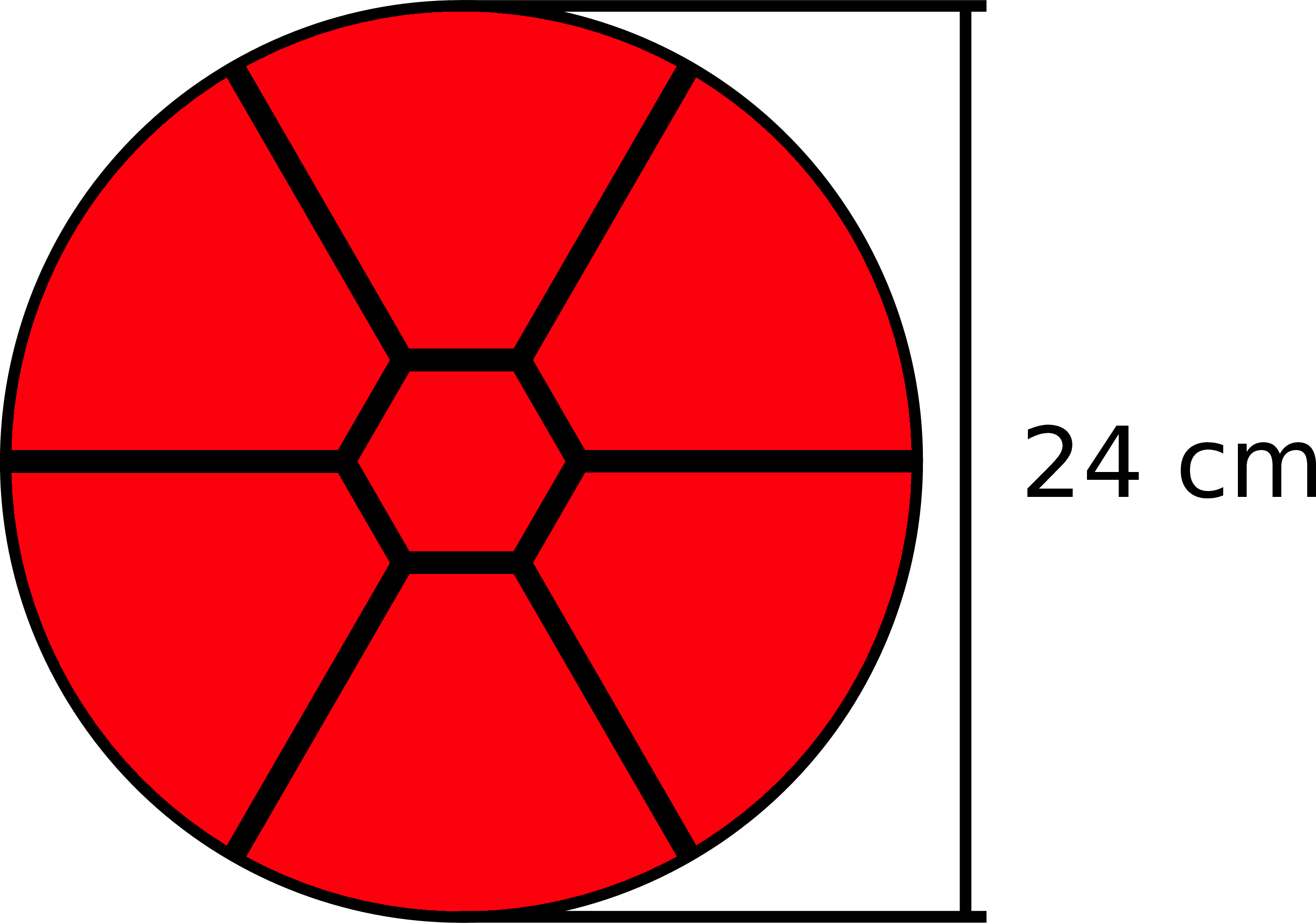}
    \caption{Possible detector design. Sunward direction is into/out of the page. A hexagonal prism of GAGG and six circle wedges of GAGG (red) fill a circular area. Space is left (black) for wires and support structure to run between volumes. Specific placement of SiPMs for this design are not finalized.}
    \label{fig:HexWindow}
\end{figure}

For a small space-based detector, other detector schemes might be able to better use the space than directly mimicking NO$\nu$A. A detector could consider small voxels in the shape of a hexagonal prism to fill the space, Figure \ref{fig:HexDetector}. Alternately, a circular piece could be divided into circular wedges with a central hexagonal region, Figure \ref{fig:HexWindow}. 

Either of these configurations carry different design considerations such as minimizing space lost to wiring and maximizing the ability to resolve particles and energy depositions within the volume. If we can optimize a design for particle identification based on track shape and length through the detector's voxels, as in experiments like NO$\nu$A \cite{ref:NOvAEventClassification}, while maintaining a large fiducial volume, a future design may be able to dramatically improve on the detection efficiency of the scientific mission.

\section{Neutrino Physics in Space}\label{sec:nuSpace}

\subsection{Disadvantages}\label{sec:nuSpace:disadvantage}
Most neutrino detectors have the luxury of being built under many tons of rock, snow, dirt, or other surface material to provide passive shielding from the energetic cosmic ray showers coming from the sky. For example, the Homestake Mine has meter water equivalent (mwe) shielding of up to 4290 mwe \cite{ref:HomestakeMWE}. A space-based detector will have to contend with the full flux of energetic cosmic gamma rays, electrons, protons, and heavier nuclei. The first level of shielding for these events comes from the outermost iron shell, as described in Section \ref{sec:Design}.

\begin{figure}[htbp]
    \centering
    \includegraphics[width=0.55 \textwidth]{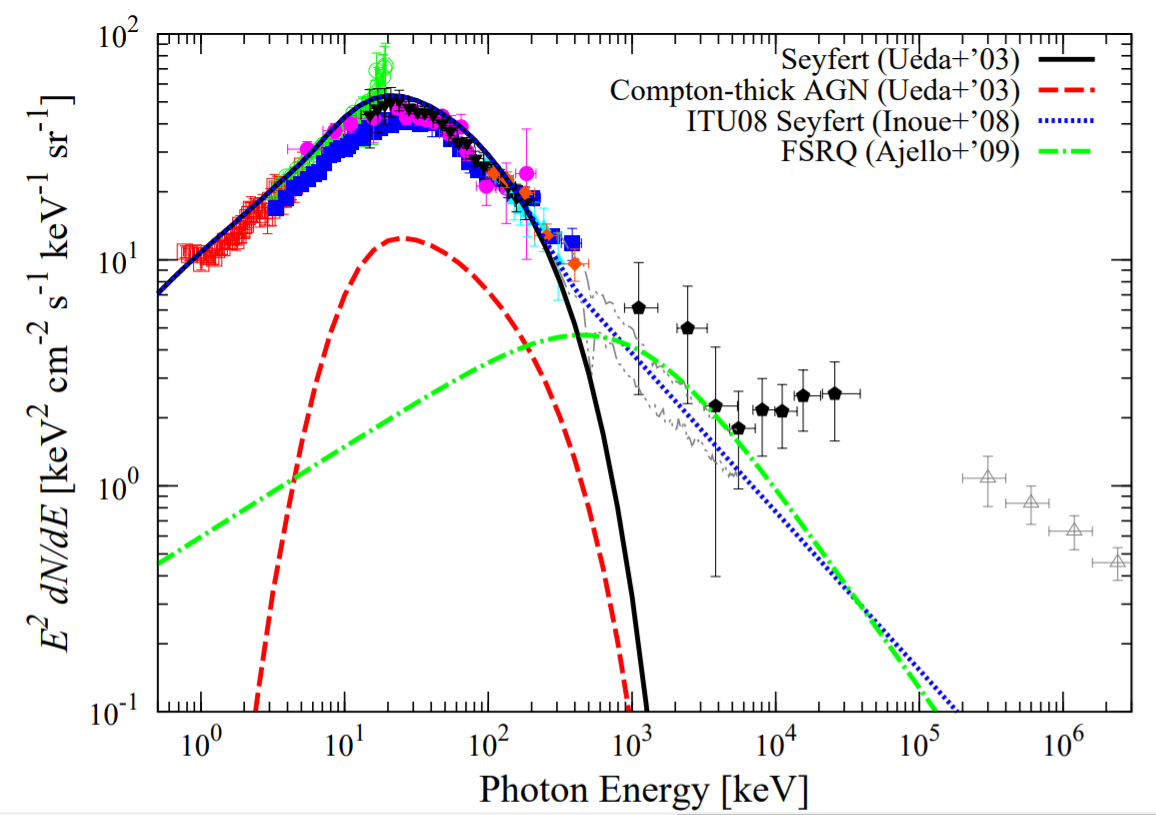}
    \caption{Cosmic gamma ray spectrum vs photon energy. Fermi gamma-ray data extends to 100 GeV, but the contribution to the total rate is small \cite{ref:Gamma_Background}.}
    \label{fig:cosmicGammaFlux}
\end{figure}

The first significant source of background events in space is the gamma ray/x-ray background. Our passive shield, 1.16 cm of iron, can stop the energy of gamma rays up to 600 keV with better than $2\cdot10^{-4}$ attenuation \cite{ref:NIST_x-ray_Attenuation}. This should be more than sufficient to shield against the gamma ray background. The total gamma background can be seen in Figure \ref{fig:cosmicGammaFlux} from 0.5 keV to 2.5 GeV. The sufficiently energetic gammas, in the energy range of 500 keV and above, create a background of approximately 130 m$^{-2}$ sr$^{-1}$ s$^{-1}$ \cite{ref:Gamma_Background}. Our electronics are being designed so that they are sufficient to deal with this noise on the order of 500 Hz. Noise from the planned SiPMs for detector readout is expected be well below the high light yield from the GAGG scintillator, and so should not affect detector operations.

\begin{figure}[htbp]
    \centering
    \includegraphics[width=0.45 \textwidth]{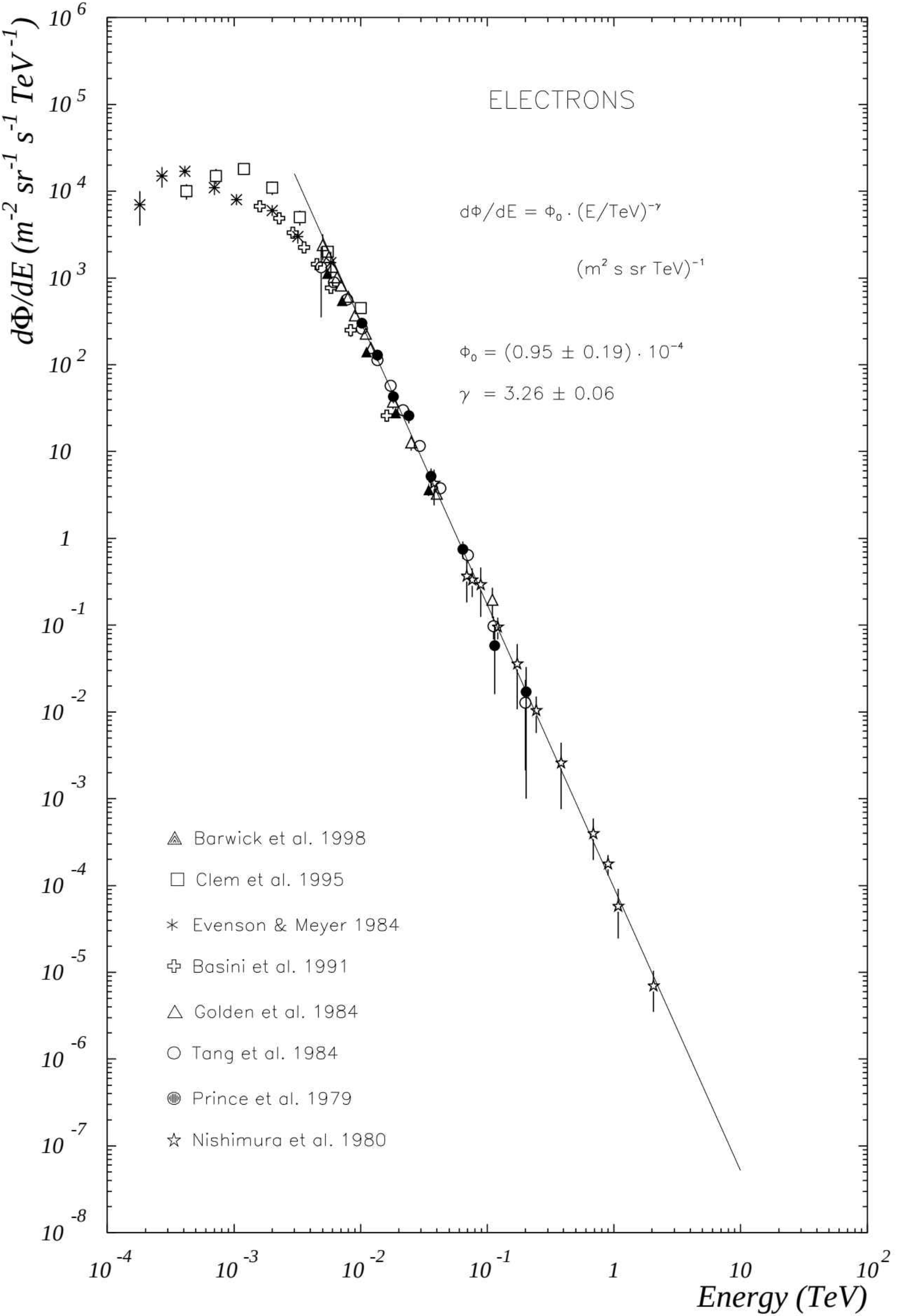}
    \caption{Cosmic electron differential flux vs electron kinetic energy \cite{ref:ElectronBackground}.}
    \label{fig:cosmicElectronFlux}
\end{figure}

There are two major sources of electrons in terms of space conditions. The solar wind electrons have a characteristic temperature of 12 eV at the Earth \cite{ref:electronTemp}. This temperature will increase, using the worst case of Boldyrev, Forest, and Egedal, to a maximum temperature of 130 eV \cite{ref:electronTemperatureRadius}. These are both well below energies at which any significant fraction of electrons might penetrate the satellite's shielding. The second source is the cosmic background electrons. These background electrons, Figure \ref{fig:cosmicElectronFlux}, have been measured with a total rate of approximately 50 m$^{-2}$ sr$^{-1}$ s$^{-1}$ on the MeV-TeV regime. The iron shield will moderate electrons and, in the continuously slowing down approximation (CSDA), the shielding will be 100\% effective at energies up to 15 MeV \cite{ref:NIST_CSDA_Ranges}. At energies beyond this, there will begin to be penetration, and we will rely on active veto rejection techniques such as topological cuts based on the pulse height and shape in the science volume or looking for time-coincident energy deposits in the vetoing volume surrounding the science volume.

Similar to electrons, the solar wind and cosmic background are the primary sources of positively charged particles. The solar wind protons have a characteristic temperature of 18 eV, and the maximum temperature measurement of the DSCOVR solar wind warning probe in the Earth-Sun L1 Lagrange point has measured maximum proton temperatures of approximately 86 eV. These energies are well below the MeV scales of the passive shielding \cite{ref:SpaceWeather}. The total cosmic flux of protons and heavier nuclei is well measured, and the total flux is on the order of 3000 m$^{-2}$ sr$^{-1}$ s$^{-1}$ from 100 MeV and up \cite{ref:PDG}. In the CSDA, the shield will be able to stop protons of energy up to 80 MeV, and it will be able to stop alpha particles up to 300 MeV.

\begin{figure}
    \centering
    \includegraphics[width=0.45\textwidth]{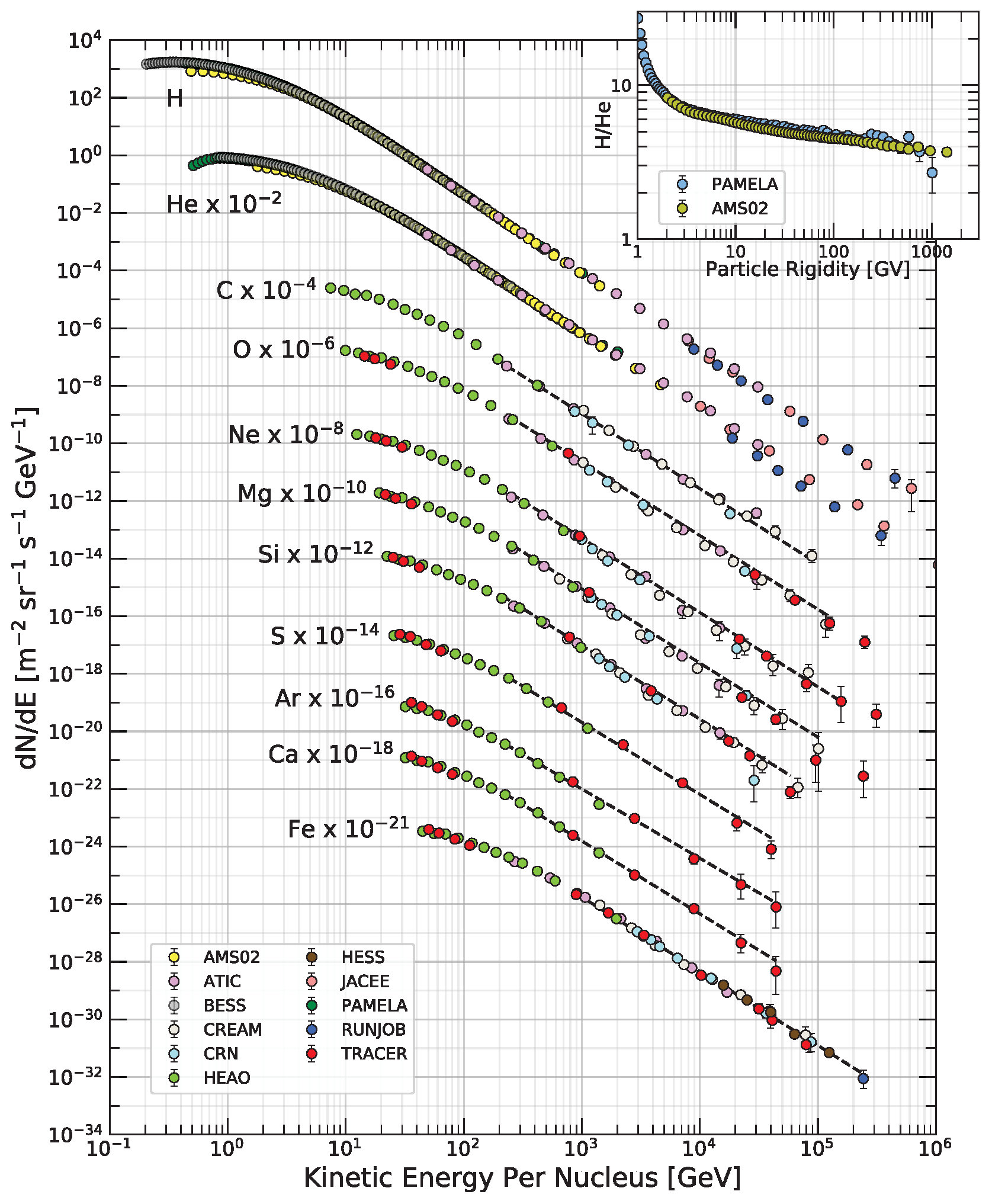}
    \caption{Fluxes of nuclei of the primary cosmic radiation in particles per energy-per-nucleus are plotted vs energy-per-nucleus. The inset shows the H/He ratio as a function of rigidity \cite{ref:PDG}.}
    \label{fig:NucleiFlux}
\end{figure}

\subsection{Advantages}\label{sec:nuSpace:advantage}
In the introduction, we touched briefly on one advantage of approaching the Sun, and here expand on the scope of what it means to go close to the Sun. The most obvious advantage is the previously mentioned inverse square scaling. Most electromagnetically interacting particles exiting the sun, such as electrons and photons, are subject to random walks through the very dense medium of the sun, and they can take thousands of years to exit. Due to the weakly interacting nature of the neutrino, we can instead expect the flux of neutrinos to fall off as an inverse square of the distance from which they were generated.

The dramatically increased flux from the inverse square scaling allows for a modestly sized detector, on the order of one hundred kilograms of active volume. For a $3R_\odot$ science mission, this would be the flux-equivalent of a 1 kTon solid scintillating crystal detector, which is comparable to large Earth-based experiments like the KamLAND, a 1 kTon mineral oil scintillator detector, and Super-K, a 50 kTon water Cherenkov detector \cite{ref:KamLAND,ref:SuperKDetector}. As we approach the Sun, the flux spectrum will change slightly due to the shape of fusion in the core, but even on the scale of a 3$R_\odot$ closest approach the variation in flux from the near peak of the fusion region to the far peak is on the order of 1\%.

\begin{figure}[htbp]
    \centering
    \includegraphics[width=0.55 \textwidth]{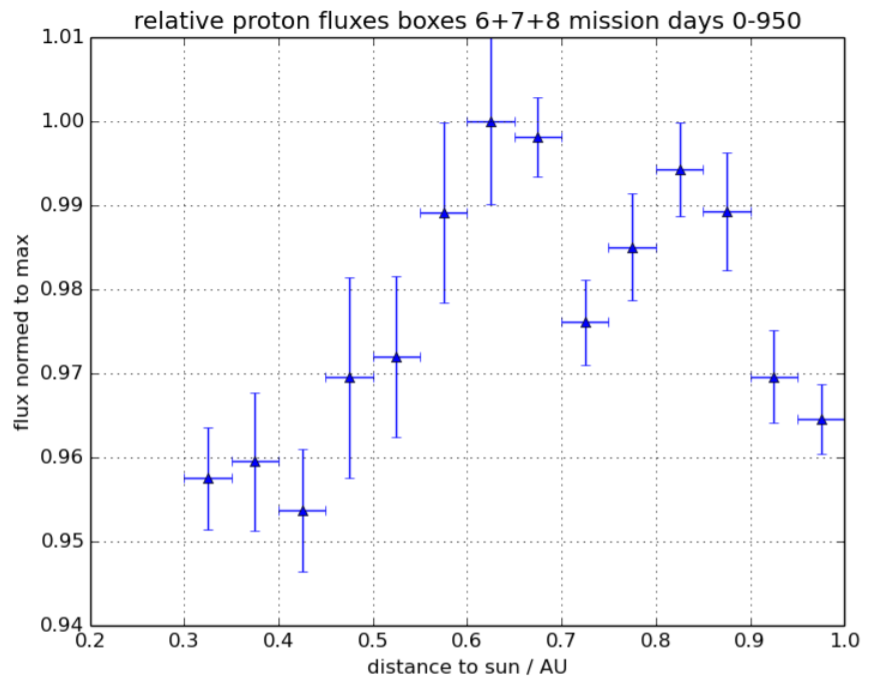}
    \caption{Flux of protons from 250-700 MeV by distance from Sun \cite{ref:HeliosNASA}}.
    \label{fig:cosmicProtonFlux}
\end{figure}

It is worth noting that there does not seem to be much risk of increased of backgrounds on approach to the Sun. The Helios mission measured the proton flux down to 0.3 AU ($4.5\cdot 10^7$ km or $64.5R_\odot$), Figure \ref{fig:cosmicProtonFlux}, and found that the flux was unexpectedly flat. This phenomenon has been explained by the modulation of the high energy cosmic particles by the Sun's magnetic field and the acceleration of the low energy solar wind as it moves from the Sun \cite{ref:CosmicModulation, ref:SolarWindAcceleration}.

Last, it is a straightforward feature of the geometry in orbit that the resolution of a detector increases with an inverse square scaling. Because of this, we get the same relative angular resolution scaling that the neutrino flux sees in Table \ref{tab:fluxTable}. If our detector is only capable of one tenth the angular resolving power of detectors like Super Kamiokande, a relatively conservative mission to 20 $R_\odot$ would have significantly improved resolving power. This sort of mission is well within our launch capabilities, and as the Parker Solar Probe (PSP) is demonstrating, cooling such a payload is within our current engineering capabilities. This detector is sufficiently similar that during a NASA Advanced Concepts Office study for the project, the engineers used a slightly smaller copy of the PSP as the heat shield for the spacecraft \cite{ref:NIACPhaseII}.

\section{Gallium Double Pulsing}\label{sec:GaDoublePulse}
In section \ref{sec:nuSpace}, we introduced the difficulty of measurement using traditional neutrino detection methods due to high background rates. To counteract this effect, we are designing a method of looking for electron-type neutrinos via a double pulse signal. The interaction of gallium with neutrinos has been of scientific interest since the early 1990's with the GALLEX experiment and the SAGE experiment \cite{ref:GALLEX,ref:SAGE}. These experiments used radiochemical methods to determine the number of neutrino interactions which had taken place. These experiments were, and our experiment currently is, based on the specific interaction of a neutrino transmuting gallium into ionized germanium, and itself into an electron, (\ref{eq:GalliumInteraction}). Due to the specifics of this interaction discussed in section \ref{sec:GaDoublePulse:Interaction}, an isotopically pure detector would be advantageous in comparison with the naturally occurring gallium. The two isotopes naturally occur as 60.1\% Gallium 71 and 39.9\% Gallium 69, and both isotopes are commercially available.

\begin{equation}\label{eq:GalliumInteraction}
    \nu_e+~_{31}^{69}\textrm{Ga}\to e^-+~_{32}^{69}\textrm{Ge}^+
\end{equation}

It is an important to note that the proposed method is not unique to any particular gallium-loading method, nor is it strictly necessary to use gallium. Several other elements could be used for solar neutrino measurements, such as indium. Our work has focused on gallium because it has the smallest invariant mass difference, $Q=235.7$ keV, and largest cross section of other double-pulse candidates. Our current designs focus on the GAGG scintillating crystal, but as we have mentioned, any gallium-loaded material could be used in this double-pulsing schemes. In particular, Gallium III Oxide is an interesting scintillator in development because of its high gallium mass fraction in comparison with other crystals. 

\subsection{Gallium-Neutrino Interaction}\label{sec:GaDoublePulse:Interaction}
We propose using a prompt scintillation method to determine when a neutrino interaction has occurred. The neutrino interaction will often leave the germanium atom without any excitation energy, but about half of the interactions will leave the germanium nucleus in an excited state, see Table \ref{tab:NeutrinoReactions}. This motivates a method to look for an initial electron energy deposit followed by a gamma ray with characteristic timing. This double pulsing allows us to reject a large fraction of events which bypass our shielding. This allows the potential for favorable signal-to-background ratios even in high-background environments like those of space.

The two isotopes of gallium present two different possibilities for neutrino detection. The first isotope that we will discuss is gallium 71. Gallium 71 is transmuted into germanium 71 during interactions with electron type neutrinos. When the neutrino is sufficiently energetic, this will release an electron alongside the final germanium nucleus. The mass difference between gallium and germanium 71 is 235.7(18) keV, and when combined with the mass of the electron, this gives a bare threshold of 746.7(18) keV for the energy of a neutrino able to eject an electron at rest from the ground state of germanium 71 \cite{ref:A=71,ref:PDG}. The gallium ground state has spin -3/2, whereas the germanium ground, first, and second excited states have spins of -1/2, -5/2, and +9/2 respectively. The -5/2 spin state has a half-life of 79 ns, and the metastable spin +9/2 state has a half life of 20.2 ms. The most likely excited state of the germanium is the -5/2 state, followed by several higher energy levels. The +9/2 state should be very unlikely in comparison \cite{ref:GalliumAnomaly,ref:GalliumLevelsTheory2021}.

As mentioned in section \ref{sec:Design:Continuing}, it is promising to design a highly-segmented detector made of small GAGG crystals. The detector will use SiPMs to read out the energy deposited in each of the GAGG cells, and this segmentation allows for sophisticated particle identification algorithms like have been used in the NOvA experiment \cite{ref:NOvAEventClassification}. A highly segmented design has not yet been implemented in simulation, but the work done in particle identification by the NOvA experiments should allow our detector to identify particle type and direction for more effective vetoing.

\begin{table}[htp!]
\begin{center}
\begin{tabular}{ | c | c | c | c | }
\hline
{Reaction Products} & {Energy Threshold} & {Photon Energy} & {Half Life} \\ 
 \hline\hline
 Ge$_{32}^{71E2}+e^-$& 0.408 MeV & 0.175 MeV & 79 ns \\  
 \hline
 Ge$_{32}^{69M1}+e^-$& 2.313 MeV & 0.086 MeV & 5 $\mu$s\\ 
 \hline Ge$_{32}^{69M2}+e^-$&2.624 MeV &0.397 MeV &2.8 $\mu$s\\ 
 \hline
\end{tabular}
\caption[Table of most likely excited states from the reaction of neutrinos with gallium, their decay products, and half-lives.]{Table of most likely excited states from the reaction of neutrinos with gallium, their decay products, and half-lives.}
\label{tab:NeutrinoReactions}
\end{center}
\end{table}

\subsection{Simulations}\label{sec:GaDoublePulse:Simulation}
To study the detector's design, Section \ref{sec:Design}, a fast Monte Carlo has been implemented to simulate the production of false double pulses that can mimic a neutrino signal. Geant4, a c++ particle physics library, has been used to model the detector to produce the inputs to the fast Monte Carlo. The model implements optical photons emitted during scintillation to produce pulses used to identify the true pulse shapes from the gallium signals and false pulse shapes from background particles, and it uses the same geometry as in Section \ref{sec:Design:SecondDesign}, Figure \ref{fig:DetctorV1}. Particles were fired isotropically from a sphere containing the detector volume. . A true event is generated by creating an isotropically emitted electron, Lorentz boosting it into the incident neutrino’s frame, and then releasing an isotropic gamma ray delayed by the excited germanium’s half life. The simulation then counts any photons incident on the photosensitive volumes. Whenever such a hit is counted, the time is recorded. If there is a sufficient time gap, 25 ns in the current model, then the photons are separated into one or more individual pulses. These pulses were analyzed to generate information about the topology of the pulses, the time at which the peak optical photon production occurs, and the pulse shape ratio. The pulse shape ratio is defined as the ratio of the area of the pulse beyond 5 nanoseconds after the peak time of the pulse. This information is then used to generate criteria for determining if a particular pulse satisfies the electron or gamma ray topology.

The veto system surrounds the detector volume with scintillating plastic, as described in Section \ref{sec:Design}. To study the proton, alpha particle, and gamma ray single rejection rate, each pulse was checked against the signal in the veto array. If the veto volume detects optical photons in a time consistent with a signal inside the main detector volume, the entire pulse corresponding to that veto signal is rejected. Any remaining pulses are analyzed based on the energy and timing pulse topologies determined by the true signals described above. Any signal that did not have a veto signal corresponding it, but also does not fit the same pulse topology as the $e^-$ + $\gamma$, is rejected as well. Any signals remaining from these criteria are treated as electron or gamma ray signal candidates. If a signal coinciding with the gamma-ray pulse topology follows within 5 half lives of a signal coinciding with the prompt-electron signal, then this paired signal is counted as a false positive neutrino event. A large number of cosmic rays, up to 2.5 million, were used at energies consistent with the cosmic ray spectrum to generate these pulses for analysis.

To study multiple coincidence of GCRs, a Geant4 simulation with this veto criteria was then repeated over a wide range of energies for cosmic ray protons, alpha particles, and gamma rays. Other sources of background noise have not been studied at present. Some, like higher atomic number galactic cosmic rays, are not expected to be difficult to reject. Other sources of noise, such as radioisotope contamination or activation by incident radiation, have not yet been studied. These background sources have been a challenge for past neutrino detectors, and may represent a significant technical hurdle for a space mission if sufficient radioisotope purity and radiation shielding cannot be achieved.

Once the pulses are characterized, the ROOT data analysis framework begins the fast Monte Carlo by reading in the pulse information for galactic cosmic rays. From that information one can determine the rate at which these pulses are generated. It is found that the rate for protons is 230.57 pulses per second, the rate for alphas is 29.00 pulses per second, and the rate for gamma rays is 9.46 pulses per second. A Poisson distribution is applied to each of these rates to generate a random number of events within each second. It is possible that multiple pulses from the same event will qualify, and if that event number from the large data run is selected, all associated pulses from that particle are added into the analysis. The process continues until one day’s worth (86,400 seconds) of simulated pulses are generated. The master array then reorders all pulses from protons, alphas, and gammas from the earliest time to the latest time.

After the re-ordering, the fast Monte Carlo starts to pair pulses together. It considers potential double pulses that occur within 3, 5, and 10 half-lives of the gallium interaction. The first pulse must qualify as a primary electron, and the next pulse must qualify as either a secondary gamma ray or the secondary decay of both gammas and electrons. The program pairs the current pulse in the loop with all successive pulses until the edge of the timing window is met, and the program then determines how many false double pulses occur in each timing window, and whether the pulse came from a proton, an alpha, or a gamma event. 

\begin{figure}[htbp]
    \centering
    \includegraphics[width=0.75 \textwidth]{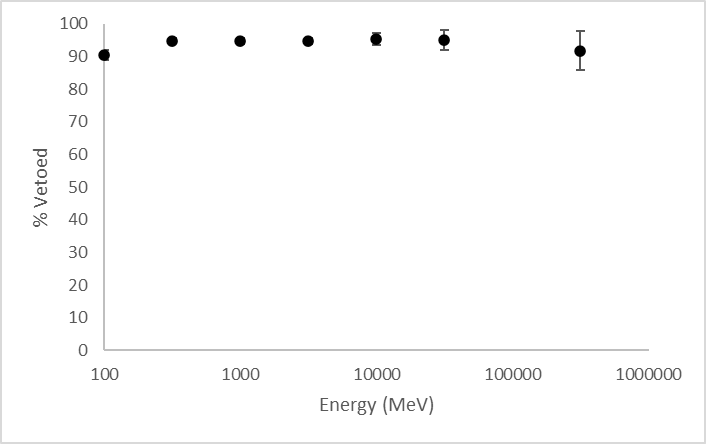}
    \caption{Simulated proton single-event rejection rate vs proton energy. Error range is $1\sigma$.}
    \label{fig:protonRejection}
\end{figure}

\begin{figure}[htbp]
    \centering
    \includegraphics[width=0.75 \textwidth]{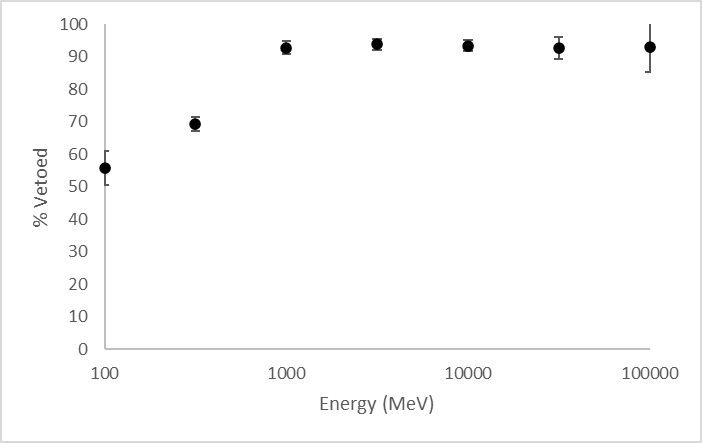}
    \caption{Simulated alpha particle single-event rejection rate vs alpha particle energy. Error range is $1\sigma$.}
    \label{fig:alphaRejection}
\end{figure}

\begin{figure}[htbp]
    \centering
    \includegraphics[width=0.75 \textwidth]{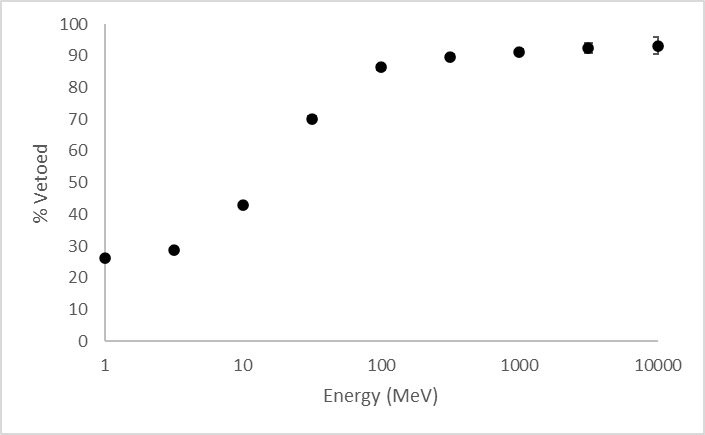}
    \caption{Simulated gamma ray single-event rejection rate vs gamma ray energy. Error range is $1\sigma$.}
    \label{fig:gammaRejection}
\end{figure}

Between the first and second designs of the detector, Section \ref{sec:Design}, we saw a marked improvement in the rejection rate of background noise. This arose primarily from the segmentation of the detector from one cylinder into four. With four detection volumes, accidental false signals now have to exhibit correlation in space as well as the energy and timing topologies. Current simulations suggest that both alpha particles and protons are easy to separate from true events, and the single-coincidence rejection rate on these is nearly 100\%. Alpha particles can be somewhat difficult to reject in the 100 - 1000 MeV energy range, as seen in Figure \ref{fig:protonRejection}, but Figure \ref{fig:alphaRejection} shows that protons are rejected well over the entire energy regime of interest for a potential solar mission. Gamma rays have been more difficult to reject, with their single-coincidence rejection rate only exceeding 80\% near 100 MeV as show in Figure \ref{fig:gammaRejection}. While they are the most difficult to completely veto individually, the gamma rays are sufficiently low rate that they do not provide the largest background in comparison to the protons and alpha particles.

We expect that these rates will continue to improve with improved detector design configurations and continued pulse topology studies. As noted earlier, a detector more highly segmented than the current four cylinder design, Sections \ref{sec:Design:SecondDesign}, \ref{sec:Design:Continuing}, may allow for much better differentiation between accidental double-pulse candidates from backgrounds by looking at the energy deposition and tracks of the background particles.

\section{Summary}\label{sec:Summary}
In summary,we propose a near-solar neutrino detector design using the well-studied transmutation of gallium into germanium. This detector will be able to take advantage of the inverse square scaling of the neutrino flux to be the mass-equivalent of a detector on the scale of kilotons while at closest solar approach. By flying in the space between the Earth and the Sun, such a detector will be capable of probing distance scales larger than the diameter of the Earth, the upper limit for beamline experiments, and the detector will be capable of probing distance scales smaller than 1 AU ($1.5\cdot10^7$ km), the smallest baseline for extraterrestrial neutrino sources.

Such a detector must contend with high-rate cosmic ray backgrounds. We propose to solve this problem by looking for a double-pulsed gallium signal. This double pulsing allows for dramatically improved rejection rates compared to single coincidence rejection. Future work will continue to refine detector pulse topology rejections. Our rejection rate in-simuo is already at 80\% for gamma rays with little pulse topology or timing cutting, and nearly 100\% for other background particles in similar regimes. Ongoing studies into detector design and pulse analysis continue to improve these rejection rates.

\section{Acknowledgements}\label{sec:Acknowledgements}
We would like to thank Nasser Barghouty of NASA Headquarters for helping Nick Solomey put together the initial idea for the $\nu$SOL project during a NASA summer fellowship.

Funding: This work was supported by the NASA NIAC Program [grant numbers 80NSSC18K0868, 80NSSC19M0971]; MSFC CAN [grant number 80MSFC18M0047]; Wichita State University MURPA; NASA EPSCoR PDG.

 \bibliographystyle{elsarticle-num} 
 \bibliography{cas-refs}





\end{document}